# CRAB WAIST COLLISION STUDIES FOR e+e- FACTORIES

M. Zobov, P. Raimondi, LNF-INFN, Frascati, Italy
D. Shatilov, BINP, Novosibirsk, Russia
K. Ohmi, KEK, Tsukuba, Japan

*Abstract*

Numerical simulations have shown that the recently proposed "crab waist" scheme of beam-beam collisions can substantially boost the luminosity of existing and future electron-positron colliders. In this paper we describe the crab waist concept and discuss potential advantages that such a scheme can provide. We also present the results of beam-beam simulations for the two currently proposed projects based on the crab waist scheme: the DAΦNE upgrade and the Super B-factory project.

## INTRODUCTION

In high luminosity colliders with standard collision schemes the key requirements to increase the luminosity are: the very small vertical beta function $\beta_y$ at the interaction point (IP); the high beam intensity I; the small vertical emittance $\varepsilon_y$ and large horizontal beam size $\sigma_x$ and horizontal emittance $\varepsilon_x$ for minimization of beam-beam effects. However, $\beta_y$ can not be much smaller than the bunch length $\sigma_z$ without incurring in the "hour-glass" effect. It is, unfortunately, very difficult to shorten the bunch in a high current ring without exciting instabilities. In turn, the beam current increase may result in high beam power losses, beam instabilities and a remarkable enhancement of the wall-plug power. These problems can be overcome with the recently proposed Crab Waist (CW) scheme of beam-beam collisions [1] where a substantial luminosity increase can be achieved without bunch length reduction and with moderate beam currents.

These advantages have triggered several collider projects exploiting the CW collision potential. In particular, the upgrade of the Φ-factory DAΦNE is aimed at increasing the collider luminosity toward to $10^{33}$ cm$^{-2}$s$^{-1}$ [2] to be compared with $1.6 \times 10^{32}$ cm$^{-2}$s$^{-1}$ obtained during the last DAΦNE run for the FINUDA experiment [3]. At present the upgraded DAΦNE is being commissioned and the first crab waist collisions are expected in the winter/spring 2008 [4]. Besides, the physics and the accelerator communities are discussing a new project of a Super B-factory with luminosity as high as $10^{36}$ cm$^{-2}$s$^{-1}$ [5], i.e. by about two orders of magnitude higher with respect to that achieved at the existing B-factories at SLAC [6] and KEK [7]. The decision on the Super B-factory construction will depend much on the results of the CW collision tests at DAΦNE.

In the following we briefly discuss the Crab Waist collision concept and present results of beam-beam simulations for the DAΦNE upgrade and for the Super B-factory project.

## CRABBED WAIST CONCEPT

The Crab Waist scheme of beam-beam collisions can substantially increase collider luminosity since it combines several potentially advantageous ideas. Let us consider two bunches with the vertical $\sigma_y$, horizontal $\sigma_x$ and longitudinal $\sigma_z$ sizes colliding under a horizontal crossing angle $\theta$ (as shown in Fig. 1a). Then, the CW principle can be explained, somewhat artificially, in the three basic steps.

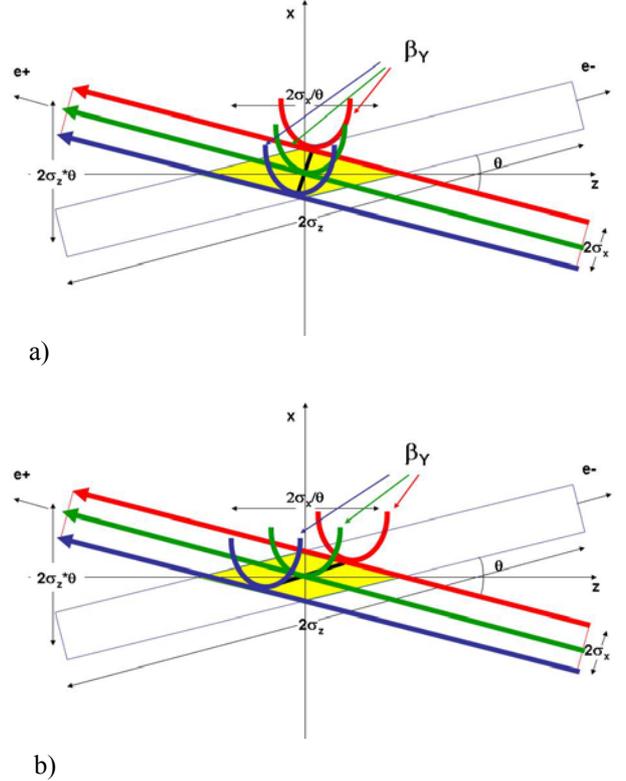

a)

b)

Fig. 1 Crab Waist collision scheme
((a) – crab sextupoles off; (b) – crab sextupoles on)

The first one is large Piwinski angle. For collisions under a crossing angle $\theta$ the luminosity $L$ and the horizontal $\xi_x$ and vertical $\xi_y$ tune shifts scale as (see, for example, [8]):

$$L \propto \frac{N\xi_y}{\beta_y^*} \propto \frac{1}{\sqrt{\beta_y^*}}; \quad \xi_y \propto \frac{N\sqrt{\beta_y^*}}{\sigma_z \theta}; \quad \xi_x \propto \frac{N}{(\sigma_z \theta)^2}$$

Here the Piwinski angle is defined as:

$$\phi = \frac{\sigma_z}{\sigma_x} tg\left(\frac{\theta}{2}\right) \approx \frac{\sigma_z}{\sigma_x}\frac{\theta}{2}$$

with $N$ being the number of particles per bunch. Here we consider the case of flat beams, small horizontal crossing angle $\theta \ll 1$ and large Piwinski angle $\phi \gg 1$.

The idea of colliding with a large Piwinski angle is not new (see, for example, [9]). It has been also proposed for hadron colliders [10, 11] to increase the bunch length and the crossing angle. In such a case, if it were possible to increase $N$ proportionally to $\sigma_z\theta$, the vertical tune shift $\xi_y$ would remain constant, while the luminosity would grow proportionally to $\sigma_z\theta$ (see the above formulae for the luminosity and tune shifts). Moreover, the horizontal tune shift $\xi_x$ drops like $1/\sigma_z\theta$. However, differently from [10, 11], in the crab waist scheme described here the Piwinski angle is increased by decreasing the horizontal beam size and increasing the crossing angle. In this way we can gain in luminosity as well, and the horizontal tune shift decreases due the larger crossing angle. But the most important effect is that the overlap area of the colliding bunches is reduced, since it is proportional to $\sigma_x/\theta$ (see Fig. 1).

Then, as the second step, the vertical beta function $\beta_y$ can be made comparable to the overlap area size (i.e. much smaller than the bunch length):

$$\beta_y^* \approx \frac{\sigma_x}{\theta} \ll \sigma_z$$

We get several advantages in this case:

- Small spot size at the IP, i.e. higher luminosity L.
- Reduction of the vertical tune shift $\xi_y$.
- Suppression of synchrobetatron resonances [12].
- Reduction of the vertical tune shift with the synchrotron oscillation amplitude [12].

Besides, there are additional advantages in such a collision scheme: there is no need to decrease the bunch length to increase the luminosity as proposed in standard upgrade plans for B- and Φ-factories [13, 14, and 15]. This will certainly helps solving the problems of HOM heating, coherent synchrotron radiation of short bunches, excessive power consumption etc. Moreover, parasitic collisions (PC) become negligible since with higher crossing angle and smaller horizontal beam size the beam separation at the PC is large in terms of $\sigma_x$.

However, large Piwinski angle itself introduces new beam-beam resonances which may strongly limit the maximum achievable tune shifts (see [16], for example). At this point the crab waist transformation enters the game boosting the luminosity. This is the third step. The transformation is described by the Hamiltonian

$$H = H_0 + \frac{1}{2\theta} x p_y^2$$

Here $H_0$ is the Hamiltonian describing particle's motion without CW; $x$ the horizontal coordinate, $p_y$ the vertical momentum. Such a transformation produces the vertical beta function rotation according to:

$$\beta_y = \beta_y^* + \frac{(s - x/\theta)^2}{\beta_y^*}$$

As it is seen in Fig. 1b, in this case the beta function waist of one beam is oriented along the central trajectory of the other one.

The crab waist transformation gives a small geometric luminosity gain due to the vertical beta function redistribution along the overlap area. It is estimated to be of the order of several percent [17]. However, the dominating effect comes from the suppression of betatron (and synchrobetatron) resonances arising (in collisions without CW) through the vertical motion modulation by the horizontal oscillations [18, 19]. In practice the CW vertical beta function rotation is provided by sextupole magnets placed on both sides of the IP in phase with the IP in the horizontal plane and at $\pi/2$ in the vertical one (as shown in Fig. 2).

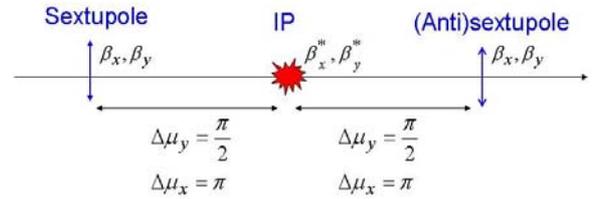

Fig. 2 Crab sextupole locations.

The crab sextupole strength should satisfy the following condition depending on the crossing angle and the beta functions at the IP and the sextupole locations:

$$K = \frac{1}{2\theta} \frac{1}{\beta_y^* \beta_y} \sqrt{\frac{\beta_x^*}{\beta_x}}$$

A numerical example of the resonance suppression is shown in Fig. 3 while beam-beam tails reduction with crab sextupoles is clearly demonstrated in Fig.7.

## DAΦNE UPGRADE SIMULATIONS

In order to estimate the achievable luminosity in DAΦNE with the crab waist scheme and to investigate distribution tails arising from beam-beam collisions, which may affect the beam lifetime, simulations with the code LIFETRAC [20] have been performed. The beam parameters used for the simulations are summarized in Table 1. For comparison, the parameters used during the last DAΦNE run with the FINUDA detector (2006-2007) are also shown [3].

As discussed above, in order to realize the crab waist scheme in DAΦNE, the Piwinski angle $\phi = \theta\sigma_x/\sigma_z$ should be increased and the beam collision area reduced: this will be achieved by increasing the crossing angle $\theta$ by a factor 2 and reducing the horizontal beam size $\sigma_x$. In this scheme the horizontal emittance $\varepsilon_x$ will be reduced by a factor of 1.7, and the horizontal beta function $\beta_x$ lowered from 1.7 to 0.2 m. Since the beam collision length decreases proportionally to $\sigma_x/\theta$, the vertical beta function $\beta_y$ can be also reduced by about a factor 3, from 1.7 cm to 0.6 cm. All other parameters will be similar to those already achieved at DAΦNE.

Table 1. Comparison of beam parameters for FINUDA run (2006-2007) and for DAΦNE upgrade

| | DAΦNE FINUDA | DAΦNE Upgrade | |
|---|---|---|---|
| $\theta_{cross}/2$ (mrad) | 12.5 | 25 | Larger Piwinski angle |
| $\varepsilon_x$ (mm×mrad) | 0.34 | 0.20 | |
| $\beta_x^*$ (cm) | 170 | 20 | |
| $\sigma_x^*$ (mm) | 0.76 | 0.20 | |
| $\Phi_{Piwinski}$ | 0.36 | 2.5 | |
| $\beta_y^*$ (cm) | 1.70 | 0.65 | Lower vertical beta |
| $\sigma_y^*$ (μm) | 5.4 (low) | 2.6 | |
| Coupling, % | 0.5 | 0.5 | Already achieved |
| $I_{bunch}$ (mA) | 13 | 13 | |
| $N_{bunch}$ | 110 | 110 | |
| $\sigma_z$ (mm) | 22 | 20 | |
| L (cm$^{-2}$s$^{-1}$) ×10$^{32}$ | 1.6 | 10 | |

Using the parameters of Table 1 and taking into account the finite crossing angle and the hourglass effect luminosity in excess of $1.0 \times 10^{33}$ cm$^{-2}$s$^{-1}$ is predicted with the achieved beam currents during the KLOE run, about 6 times higher than the one obtained until now. The only parameter that seems to be critical for a low energy machine is the high vertical tune shift: $\xi_y = 0.08$, to be compared with the value of 0.03 so far obtained at DAΦNE. In order to check whether these tune shifts (and the luminosity) are achievable we have performed the luminosity tune scans. Figure 3 shows 2D luminosity contour plots in the tune plane for the crabbed waist collisions with the crabbing sextupoles on (left) and off (right), for comparison.

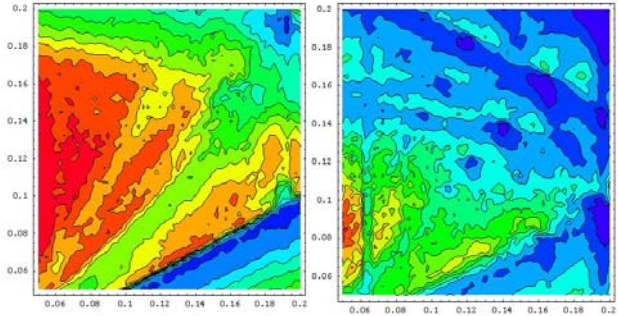

Fig.3 Luminosity tune scan ($\nu_x$ and $\nu_y$ from 0.05 to 0.20). CW sextupoles on (left), CW sextupoles off (right).

"Geographic map" colors are used to produce the plots: the brighter red colors correspond to higher luminosities (mountains), while the blue colors are used for the lowest ones (rivers and oceans). For each plot 10 contour lines between the maximum and minimum luminosities are drawn. Comparing the two plots in Fig. 3 one can see that the good luminosity region with crabbing sextupoles on is much wider than with sextupoles off since many more betatron resonances arise without CW. The absolute luminosity values are higher in the crabbed waist collisions: a peak luminosity of $2.97 \times 10^{33}$ cm$^{-2}$ s$^{-1}$ is foreseen against $L_{max} = 1.74 \times 10^{33}$ cm$^{-2}$s$^{-1}$ in the case without CW. It should be noted that the worst luminosity value obtained with CW ($2.5 \times 10^{32}$ cm$^{-2}$s$^{-1}$) is still higher than the present luminosity record at DAΦNE. Without CW the lowest luminosity value drops by an order of magnitude, down to $L_{min} = 2.78 \times 10^{31}$ cm$^{-2}$s$^{-1}$.

Strong-strong beam-beam simulations for DAΦNE upgrade have been carried out with 3D code BBSS [16]. In Fig. 4 one can see the single bunch luminosity as a function of number of turns.

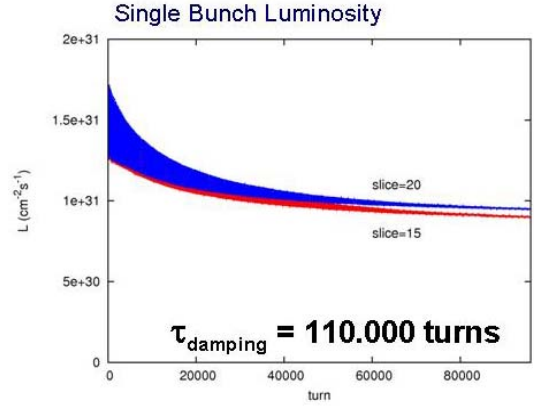

Fig. 4. Luminosity evolution in strong-strong simulations.

The simulations are very much CPU time consuming due to a large number of longitudinal slices required to simulate the crab waist conditions with the vertical beta function smaller than the bunch length. For this reason beam size and luminosity have been tracked over only one damping time. However, already from this picture one can conclude that theoretically the luminosity as high as $10^{33}$ cm$^{-2}$s$^{-1}$ (considering 110 bunches circulating in DAΦNE) is achievable and no harmful collective effects like flip-flop or coherent oscillations should be expected.

## SUPERB BEAM-BEAM SIMULATIONS

Beam-beam studies for SuperB started with a beam parameters set similar to that of the ILC damping ring (see Table 2).

Table 2. Parameters for early ILC-like design and current SuperB design. For the SuperB, the first entry is for LER and the bracketed numbers are for HER

| Parameters | ILC-like | SuperB |
|---|---|---|
| $\varepsilon_x$ (nm-rad) | 0.8 | 1.6 |
| $\varepsilon_y$ (pm-rad) | 2 | 4 |
| $\beta_x$ (mm) | 9 | 20 |
| $\beta_y$ (mm) | 0.08 | 0.30 |
| $\sigma_x$ (μm) | 2.67 | 5.66 |
| $\sigma_y$ (nm) | 12.6 | 35 |
| $\sigma_z$ (mm) | 6 | 6 |
| $\sigma_e$ (×10$^{-4}$) | 10 | 8.4 (9.0) |
| $\theta$ (mrad) | 2×25 | 2×17 |
| $N_{part}$/bunch (×10$^{10}$) | 2.5 | 6.2 (3.5) |
| $N_{bunch}$ | 6000 | 1733 |
| Circumference (m) | 3000 | 2250 |
| Damping time $\tau_s$ (ms) | 10 | 16 |
| RF frequency (MHz) | 600 | 476 |

Numerical simulations with LIFETRAC have shown that the design luminosity of $10^{36}$ cm$^{-2}$s$^{-1}$ is achieved already with 2-2.5x$10^{10}$ particles per bunch. According to the simulations, for this bunch population the beam-beam tune shift is well below the maximum achievable value. Indeed, as one can see in Fig.5, the luminosity grows quadratically with the bunch intensity till about 7.5x$10^{10}$ particles per bunch. We have used this safety margin to significantly relax and optimize many critical parameters, including damping time, crossing angle, number of bunches, bunch length, bunch currents, emittances, beta functions and coupling, while maintaining the design luminosity of $10^{36}$ cm$^{-2}$s$^{-1}$. The optimized set of beam parameters used in simulations is shown in the second column of Table 2. The most recent set of SuperB parameters can be found in [21].

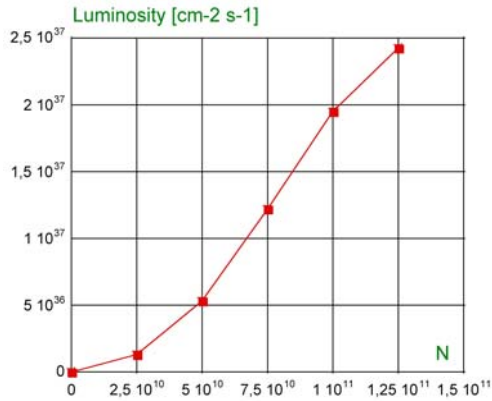

Fig. 5 SuperB luminosity versus bunch intensity

In order to define how large is the "safe" area with the design luminosity, a luminosity tune scan has been performed for tunes above the half integers, which is typical for the operating B-factories. The resulting 2D contour plot is shown in Fig.6. Individual contours differ by 10% in luminosity. The maximum luminosity found inside the scanned area is 1.21x$10^{36}$ cm$^{-2}$s$^{-1}$, while the minimum one is as low as 2.25x$10^{34}$ cm$^{-2}$s$^{-1}$. We can conclude that the design luminosity can be obtained over a wide tune area.

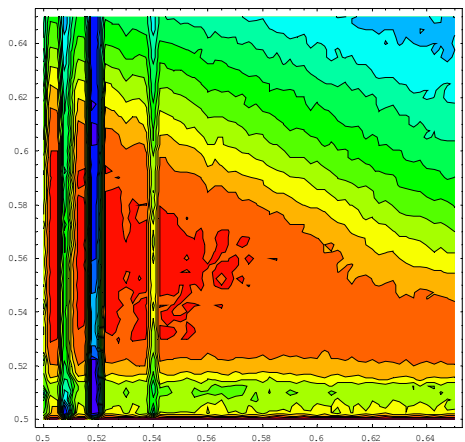

Fig. 6 SuperB luminosity tune scan (horizontal axis - $\nu_x$ from 0.5 to 0.65; vertical axis – $\nu_y$ from 0.5 to 0.65)

It has also been found numerically that for the best working points the distribution tails growth is negligible. In particular, in Fig. 7 we show distribution tails induced by the beam-beam interaction in the space of normalised betatron amplitudes as a functions of the bunch current. The unit current corresponds to the nominal bunch current, while the numbers under the pictures indicate the vertical size blow up factor – $\sigma_y/\sigma_{y0}$. As it is clearly seen comparing the last two pictures in Fig. 7, the crab sextupoles strongly suppress both the distribution tails and the vertical size blow up.

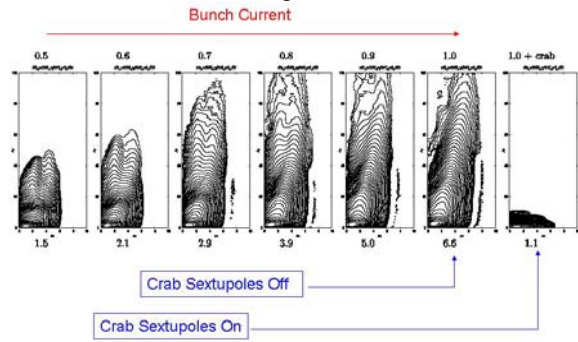

Fig. 7. Beam-beam induced tail growth as a function of bunch current.

## CONCLUSIONS

Our studies indicate that by exploiting the crab waist scheme of beam-beam collisions the luminosity of the Φ-factory DAΦNE can be pushed beyond $10^{33}$ cm$^{-2}$s$^{-1}$ level, while the luminosity of the low emittance Super B-factory can be as high as $10^{36}$ cm$^{-2}$s$^{-1}$.